# Surface Plasmon Wave plates

Amir Djalalian-Assl[1,(a)]

*[1] 51 Golf View Drive, Craigieburn, VIC 3064, Australia*

Here I investigate both numerically and experimentally, the polarization conversion capabilities of a rectangular array of holes with two unequal orthogonal periodicities. We show that it is possible to tune the periodicities in such a way that the transmitted light is circularly polarized for a nominated wavelength, $\lambda_{\mathrm{CPL}}$, when the structure is illuminated with appropriately oriented linearly polarized light at normal incidence. A device was fabricated and experiments confirmed that a degree of circular polarization of 0.89 could be achieved at the resonant wavelength.

There is increasing interest in harnessing plasmonic effects resulting from coherent electron oscillations on a metallic/dielectric interface for a broad range of applications. In particular, there are considerable prospects for replacing bulky optical components with nanophotonic analogues with thicknesses of the order of tens of nanometers [1]. The ability to manipulate the polarization of light with miniaturized optical components is highly desirable for a variety of applications [2]. For example a controllable active display using light emitting diodes has been integrated with plasmonic polarizers consisting of an array of rectangular apertures [3]. Controlling the polarization state of a Single Photon Source (SPS) provides a mechanism for defining the computational basis states [4-6]. Previously, asymmetric plasmonic cavities, such as arrays of asymmetric cross-shaped holes in a metallic film that acts as a quarter wave plate[7,8] have been demonstrated. Polarization conversion associated with arrays of L-shaped apertures in metallic films [9,10] was previously explored as was a theoretical investigation of the polarization response of an array of stereo nanoholes[11]. Furthermore, the polarization performance of a single elliptical cavity in a metal film in conjunction with periodic elliptical corrugations surrounding it, and a bullseye structure with an asymmetric cross-shaped aperture have been reported [12-14]. A periodic array of elliptical apertures in a metal film and their impact on the polarization of the transmitted light has also been investigated [15] along with a number of demonstrations of polarizing devices based on asymmetric nanoscale metallic particles [16-19]. These devices have all been based on localized resonances of asymmetric particles. Here we demonstrate, computationally and experimentally, a quarter wave plate based on surface plasmon polaritons (SPPs) propagating on the surface of a metallic film perforated with an asymmetric, rectangular arrangement of (symmetric) circular apertures. The birefringence of the structure is derived from the periodic arrangement of the apertures rather than the geometry of the apertures themselves.

To produce circularly polarized light on transmission through an array of apertures in a metal film, two orthogonal modes of equal amplitude with a phase difference of $\pi/2$ are needed. It has been shown that a biaxial nanohole array possesses two distinct orthogonal modes [20]. Here we propose, and experimentally test, a design for a plasmonic polarizer using a rectangular hole array with two unequal orthogonal periodicities in the $x$ and the $y$ directions, tuned in such a way that circularly polarized light is produced at a design wavelength when illuminated with linearly polarized light at normal incidence.

The proposed device consists of a periodic array of circular holes in a silver film, sandwiched between two semi-infinite dielectric slabs with (real) permittivities of $\varepsilon_1$ and $\varepsilon_2$. The dielectric filling the holes has a permittivity of $\varepsilon_3$.

The optical transmission associated with hole arrays of this kind can be described in terms of resonant excitation of SPP modes and Wood's anomalies [21]. It is well-known that in the case of periodic structures, such as hole arrays, standing waves arise in the steady state, which are called, SPP Bloch waves. The momentum matching condition in 2D periodic structures is given by:

$$\vec{k}_{\parallel} + i\vec{G}_x + j\vec{G}_y = \vec{K}_{SPP} \quad , \tag{1}$$

Where $\vec{k}_{\parallel}$ is the component of the incident wavevector parallel to the surface of the metal film, $G_x = 2\pi/P_x$ and $G_y = 2\pi/P_y$ are the reciprocal lattice vectors of a rectangular array with periodicities of $P_x$ and $P_y$, and $i$ and $j$ are integers.

SPP waves at the boundary between semi-infinite slabs of dielectric and metal follow conservation of momentum [22]:

$$K_{SPP} = Re\left(\sqrt{\frac{\varepsilon_{metal}\varepsilon_{diel}}{\varepsilon_{metal} + \varepsilon_{diel}}}\frac{2\pi}{\lambda_0}\right) \quad , \tag{2}$$

where $\varepsilon_{diel}$ and $\varepsilon_{metal}$ are the permittivities of the dielectric and metal respectively and $\lambda_0$ is the free-space wavelength.

For a square lattice with periodicity, $P$, the relationship between SPP Bloch modes and the lattice periodicity is given by:

$$P = \frac{2\pi}{K_{SPP}}\sqrt{i^2 + j^2} \quad , \tag{3}$$

SPP fields propagate away from a single hole as a non-uniform cylindrical wave [21] and the phase of the wave at a given point on the metal surface depends on the distance travelled. Fields are excited within the apertures which then act as scattering points on the surface. If the separations between the holes along each direction in a metallic film lead to a $\pi/2$ phase difference between the SPPs propagating in each orthogonal direction on the surface, we would then anticipate that the orthogonal components of the transmitted field would exhibit a similar phase difference. Hence, by adjusting the illumination to ensure that the amplitudes of the two components are equivalent, the film acts as a quarter wave plate. The azimuthal variation of the amplitude of the SPP field propagating away from a single hole depends on the field within the apertures which can be controlled by the orientation of the incident electric field. Hence, the direction in which the SPPs are preferentially launched is established by the component of the incident electric field parallel to the surface of the metal. The polarization angle of a normally incident plane wave, therefore, controls the amplitude of the SPPs propagating along the *x*- and *y*-directions and can be used to tune their relative strengths.

We introduce a detuning between the square lattice periodicities, *P*, in the *x* and the *y* directions associated with the SPP resonance condition at a particular design wavelength. The detuning creates a phase difference of $\pi/2$ between the two orthogonal oscillations, i.e. $\Phi_{SPP,x} - \Phi_{SPP,y} = \pi/2$ . Here $\Phi_{SPP,x,y}$ represents the relative phases of the two SPP waves along the *x*- and *y*-directions. Equations (1)-(3) do not take into account details such as the influence of the hole geometry, film thickness and the interaction between the incident field and the surface modes. They, therefore, cannot precisely predict the transmission maxima and minima associated with metallic hole arrays. For these reasons, we performed full-field electromagnetic calculations using the Finite Element Method (FEM), implemented in COMSOL Multiphysics version 4.3b, to fine-tune the device. Specifically we modeled the interaction of a normally incident, linearly polarized plane wave with an infinite array of circular holes in a thin film of silver lying parallel to the *x*-*y* plane. The upper boundary on the exit side was terminated with a perfectly matched layer (PML) to eliminate back reflection of the diffracted wave. To reduce direct transmission through the film and apertures, we set the thickness of the silver film to $h$ = 100 nm and the diameter of the hole to $d$ = 200 nm. The silver film was assumed to be supported by a glass substrate with a refractive index of $n_1$ = 1.52. Refractive index data for silver taken from Palik [23] was used in the model. In all cases, the normalized transmission $P_t/P_0$ was calculated, where $P_t$ and $P_0$ are the transmitted power through the device and through the glass substrate in the absence of the device, respectively. $P_t$ and $P_0$ were calculated by integrating the *z*-component of the transmitted Poynting vector over the area covered by one unit cell within the array.

The simulated transmission through a square array of apertures as a function of the periodicity of the array when the structure is normally illuminated with a light polarized at $\alpha = 0°$ and at a fixed wavelength $\lambda$ = 700 nm, produced a transmission maximum at a periodicity of $P$ = 394 nm. A parametric sweep over $P_y$, while setting $P_x$ = 394 nm, changes the relative phase difference between the *x* and the *y* components of the transmitted electric field, (Figure 1(a), line in red). To achieve a phase difference of 90°, two orthogonal lattice constants symmetric about $P$ = 394 nm, (i.e. $P_x \approx$ 368 nm and $P_y \approx$ 407 nm), were chosen. The blue line in Figure 1(a) depicts the normalized transmission as a function of $P_y$ for $P_x$ = 394 nm at a wavelength of 700 nm.

Note that, as foreshadowed, there is a difference between these results and those found previously for the central periodicity, $P$ = 433 nm and the corresponding length differences, $\pm\Delta P, = \pm(\pi/4)/K_{SPP} = \pm$ 54 nm found using eq. (1) - (3) for the SPP Bloch mode $(1,0)_{glass}$, at the glass/silver interface. In addition to meeting the phase requirements, to produce circularly, rather than elliptically, polarized light, the $x$ and the $y$ components of the transmitted electric field must have equal amplitudes. The SPP-incident field coupling strength in the $x$ and the $y$ directions may be controlled by varying the polarization angle of the incident light. Transmission through a rectangular array of holes with $P_x$ = 368 nm and $P_y$ = 407 nm was numerically modeled with normally incident light at various polarization angles, $\alpha$. Normalized Stokes parameters were calculated using $S_{1,2,3}/S_0$, (Figure 1(b)). The optimum incident polarization for producing circularly polarized transmitted light at $\lambda$ = 700 nm was found to be at an angle $\alpha$ = 47° from the $x$-axis. Transmission spectra , when the rectangular array was illuminated with incident polarizations $\alpha$ = {0°, 47° and 90°}, are depicted in Figure 1(c). Transmission spectra for the 0° and 90° incident polarizations show two maxima at $\lambda_x$ = 665 nm and $\lambda_y$ = 720 nm associated with the two orthogonal periodicities $P_x$ and $P_y$ respectively. Note that the curves intercept each other at $\lambda$ = 700 nm, confirming the equal amplitude requirement at the design wavelength. At 47° incident polarization, transmitted Stokes parameters were calculated, confirming transmission of CPL at $\lambda$ = 700 nm, (Figure 1(d)). Note further that at $\lambda \approx$ 663 nm, the transmitted light is linearly polarized in the $x$ direction, i.e. $S_1$ = 1. The origin of such dichroic behavior lies in the resonant transmission in the $x$ direction coinciding with the transmission suppression in the $y$ direction at $\lambda \approx$ 663 nm.

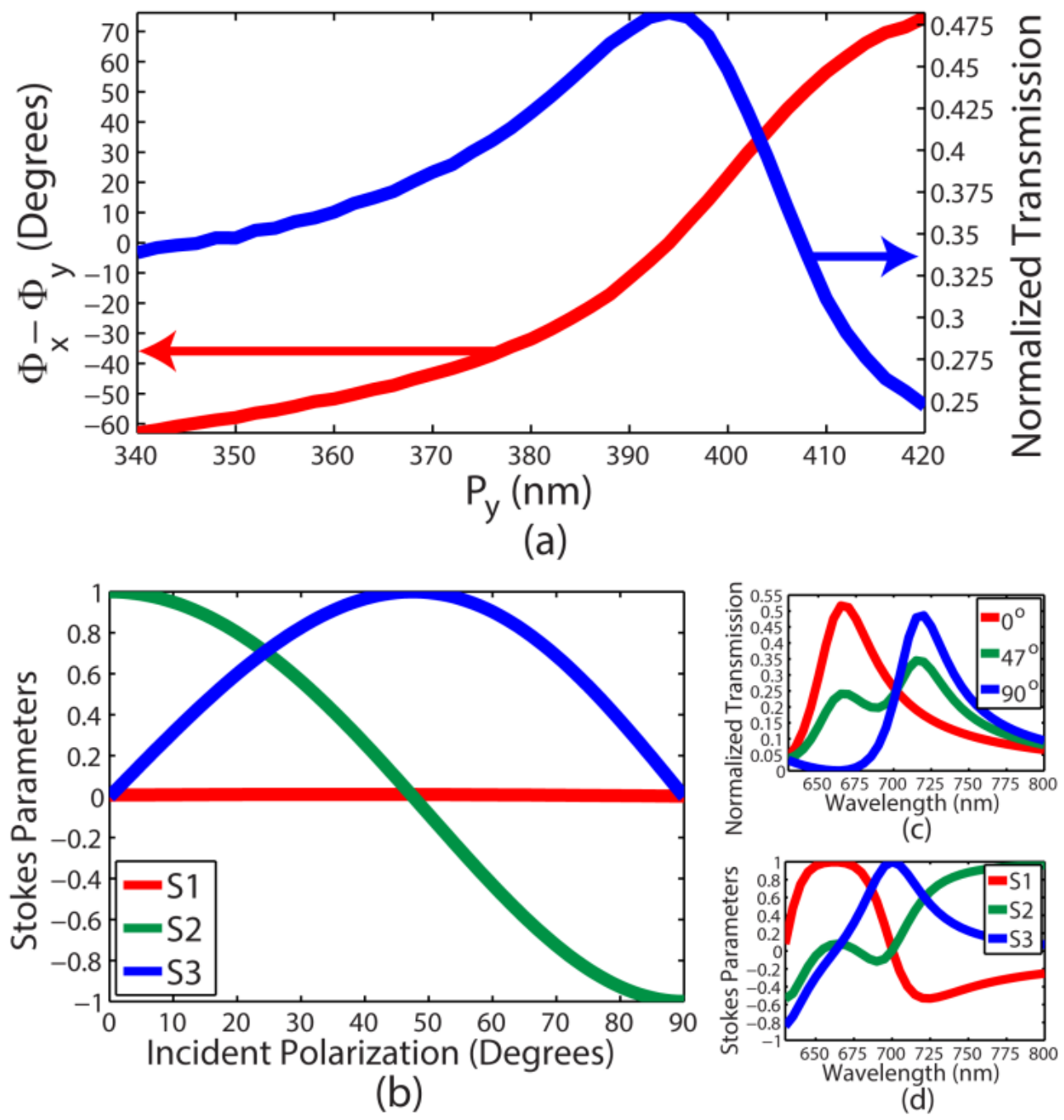

FIG 1. (a) (Line in red) The relative phase differences between the *x* and *y* components of the transmitted electric field and (Line in blue) The normalized transmission as a function of $P_y$ for $P_x$ = 394 nm at a wavelength of 700 nm. (b)-(d) Spectra of a rectangular hole array having periodicities $P_x$ = 368 nm and $P_y$ = 407 nm, (b) Normalized Stokes parameters vs. the incident polarizations. (c) Normalized intensities vs. the wavelength for incident polarizations 0°, 47° and 90°. (d) Stokes parameters vs. the wavelength for incident polarization 47°.

A top view of the simulated surface charge density on the silver/glass interface, when the device was normally illuminated from the air side with $\alpha$ = 47° at $\lambda$ = 700 nm, is depicted in Figure 2. The surface charge density and the transmitted electric field vector (represented by the red arrow) 70 nm from the glass/silver interface, were produced at $t$ = {0, $T$/8, $T$/4, 3$T$/8}, where $T$, is the period of the optical wave. The expected rotation in the surface charge density and electric field can be seen.

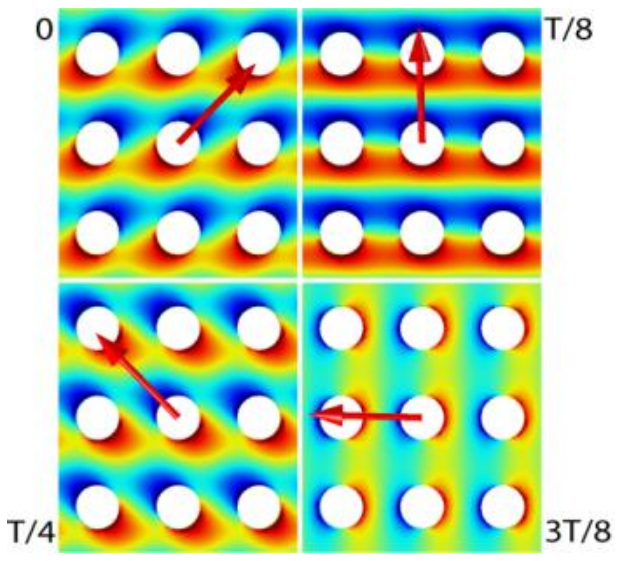

FIG 2. Surface charge density and the transmitted electric field vector (70 nm from the glass/silver interface, represented by the red arrow), were calculated at $t = \{0, T/8, T/4, 3T/8\}$, where $T$, is the period.

The results shown above are based on optimizing the performance at a design wavelength of $\lambda = 700$ nm. Simulations calculating the Stokes parameters as a function of wavelength with varying incident polarization angle, $\alpha$, provide a clearer picture of the device performance. Figure 3(a) shows that a high degree of transmitted CPL, i.e. $S_3 \approx 1$, is achievable for $40° \leq \alpha \leq 80°$, however the wavelength associated with the $S_3 \approx 1$, experiences a blue shift from 704 nm to 676 nm respectively.

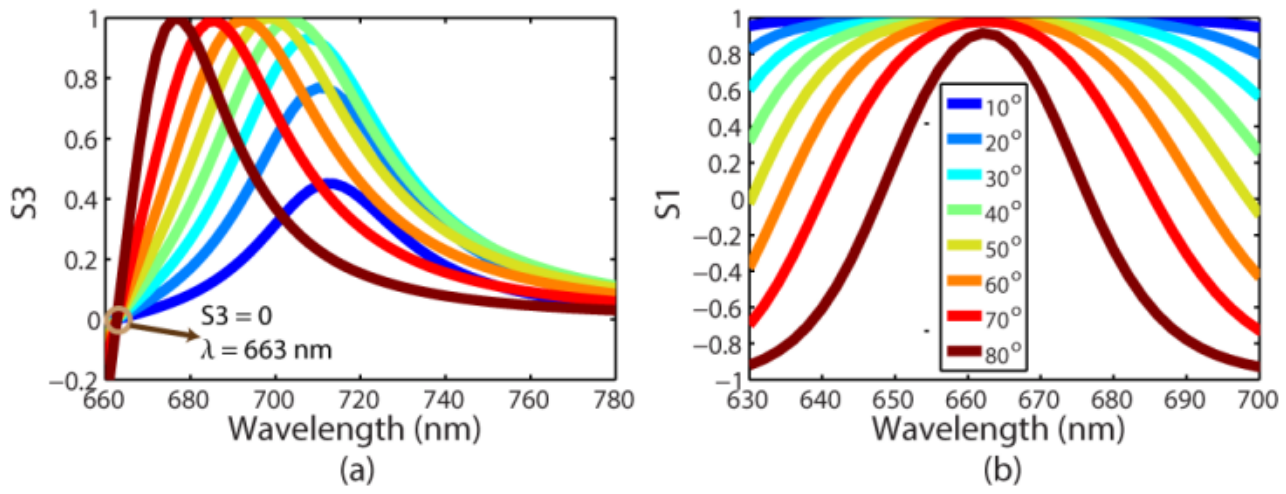

FIG 3. Simulated (a) transmitted $S_3$ and (b) transmitted $S_1$ parameters vs. the wavelength for incident polarizations $10° \leq \alpha \leq 80°$ range.

To experimentally confirm the performance of the device, a 2 nm thick germanium film (as an adhesion layer) was first deposited on a glass substrate followed by a 100 nm thick silver film and a 10 nm $SiO_2$ protective layer, using an IntlVac Nanochrome II electron beam evaporator . A rectangular array of holes with a target diameter of $d = 200$ nm and with design periodicities of $P_x = 368$ nm and $P_y = 407$ nm was milled using a Helios NanoLab 600 Focused Ion Beam (FIB). SEM images showing the fabricated structure are provided in Figure 4. The fabricated structures had periodicities of $\overline{P}_x \pm \Delta P_x = 366 \pm 13$ nm and $\overline{P}_y \pm \Delta P_y = 417 \pm 16$ nm with the array being slightly skew (< 1°). Averaging was performed over 20 periods and uncertainties are standard deviations. The average hole diameter measured $d = 181 \pm 11$ nm with some irregularities in geometry. Visually inspecting the cross-sectional SEM image we identified the $SiO_2$, silver and silver oxide layers. The silver and the silver oxide layers varied in thickness across the film and measured approximately 80 nm and 20 nm respectively.

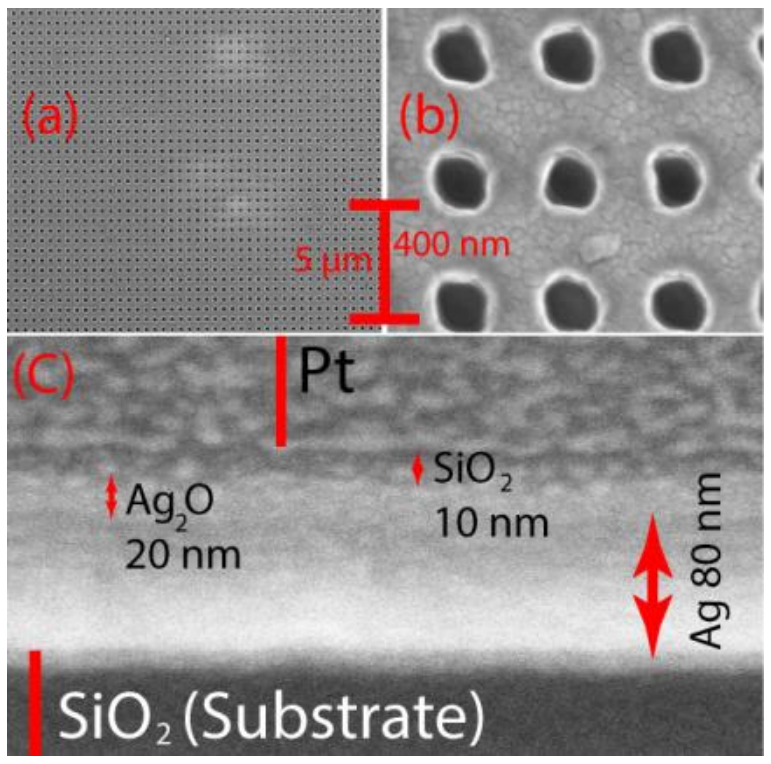

FIG 4. (a) Top view SEM image of the milled device. (b) Close-up SEM image showing the hole apertures. (c) Cross-sectional SEM image from the film.

To measure the transmission, the device was illuminated (at normal incidence) from the air/silver side with an incandescent light passed through a linear polarizer (ThorLabs LPVIS050-MP) and a Nikon C-C Achromatic condenser operating in collimator mode to produce a normally incident beam. Transmitted light was collected with a Nikon 40× objective (NA = 0.6) and analyzed with a spectrometer (Andor iDus DU920P-BR-DD CCD, controlled by Andor Solis 4.21). The transmitted intensities (normalized to those through the substrate) were measured with the incident polarization varying from 0° to 90° (where 0° corresponds to polarization in the *x*-direction). To measure the Stokes parameters, the techniques described in [24] were employed where the transmitted light was passed through a circular polarizer constructed using a linear polarizer (ThorLabs LPVIS050-MP) and a quarter waveplate (ThorLabs AQWP05M-600).

Experimental results are shown in Figure 5(a)-(d). The maximum value of $S_3$ was found to occur for an incident polarization of 43° at a wavelength of $\lambda \approx 739$ nm when $S_3 = 0.89$. The results of Figure 5 demonstrate the high degree of polarization control obtainable using this strategy. Figure 5(d) depicts the Stokes parameters and degree of polarization (DOP) when the device was illuminated with unpolarized light at normal incidence. It can be seen that the degree of linear polarization in the range $660 < \lambda < 700$ nm was found to be $0.7 < S_1 < 0.75$.

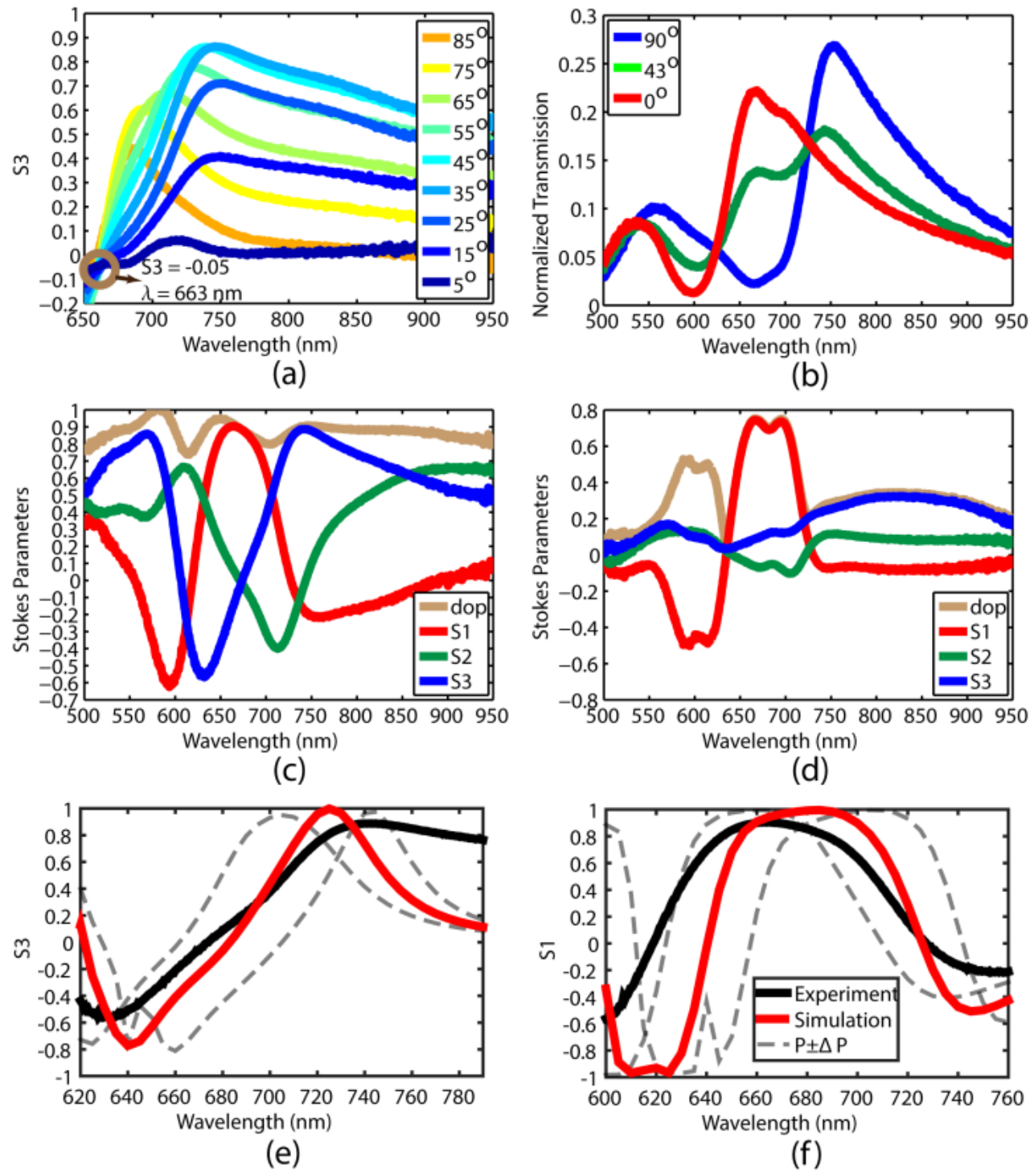

FIG 5. Experimental data showing transmission through the device. (a) Transmitted $S_3$ spectra for incident polarizations $5° \leq \alpha \leq 85°$. (b) Transmission spectra when incident field is at polarizations 90°, 43° and 0°. (c) Transmitted Stokes parameters vs. the wavelength for incident polarization 43°. (d) Transmitted Stokes parameters vs. the wavelength for un-polarized normally incident light. (e)-(f) Simulated $S_3$ and $S_1$ parameters. (Red lines) simulated results performed for the fabricated geometries. (Black lines) experimental results as in (c). Dashed lines show calculations of devices assuming the extreme range of geometric uncertainties.

The discrepancies between our predicted and experimental results are attributed to fabrication artifacts and the use of refractive index data for bulk silver in modeling nanoscale features [25]. To demonstrate the impact of fabrication artifacts on the $S_1$ and $S_3$ parameters, simulations were performed at an incident polarization angle of 43° for the fabricated geometries, although assuming a perfectly rectangular array for simplicity. Also the 20 nm $Ag_2O$ and the 10 nm $SiO_2$ were incorporated into the model. The refractive index for the $Ag_2O$ layer was set to $n$ = 2.4 [26,27]. In Figure 5(e)-(f), the dotted lines show the results of simulations performed at $(\bar{P}_x - \Delta P_x, \bar{P}_y - \Delta P_y)$ and $(\bar{P}_x + \Delta P_x, \bar{P}_y + \Delta P_y)$ giving an indication of the variability in performance associated with fabrication errors. It is apparent that fabrication uncertainties play a key role in the performance of devices.

In conclusion, I have shown that the periodicities of a rectangular array of holes perforated in metal can be tuned to produce an ultracompact, planar quarter wave plate for use at a specific wavelength. I have experimentally demonstrated a prototype device and future research should be directed at developing improved, scalable fabrication strategies to improve structure performance and predictability. Devices such as that presented here may play a role in future optical communication and sensing systems as well as in novel displays.

**Conflicts of Interest:**

I wishes to state that no other individual, other than myself (S. Amir H. Djalalian-Assl), played any role in the inception of the idea, the design of the study; in the collection, analyses, or interpretation of data; in the writing of the manuscript, and in the decision to publish the results.

**Historical Fact:**

It has come to my attention that some at the University of Melbourne have uploaded my copyrighted manuscript into the Minerva repository, without my prior knowledge or permission, associating it to some funding scheme unjustifiably.

University Library

MINERVA ACCESS

Login

**A gateway to Melbourne's research publications**

**Minerva Access is the University's Institutional Repository.** It aims to collect, preserve, and showcase the intellectual output of staff and students of the University of Melbourne for a global audience.

Minerva Access → Science → School of Physics → School of Physics - Research Publications → View Item

Search

Search the Repository

This Collection

## Surface plasmon wave plates

**Download**

Published version (2.040Mb)

Show Statistical Information

**Citations**

| Scopus | Web of Science | Altmetric |
|---|---|---|
| 8 | 8 | |

**Author**

**Document Type**

Journal Article

**Citations**

Djalalian-Assl, A; Cadusch, JJ; Teo, ZQ; Davis, TJ; Roberts, A, Surface plasmon wave plates, APPLIED PHYSICS LETTERS, 2015, 106 (4)

**Access Status**

**Open Access**

**URI**

http://hdl.handle.net/11343/116324

**DOI**

10.1063/1.4906596

**ARC Grant code**

ARC/DP110100221

Minerva Access

Depositing Your Work *(for University of Melbourne Staff and Students)*

News

FAQs

Browse

Communities & Collections

By Issue Date

Authors

Titles

grant

# Plasmonic nano-antennas for next-generation photon sources [ 2011 - 2013 ]

**Funded by** Australian Research Council
**Managed by** University of Melbourne
**Provided by** Australian Research Council

## Research Grant [Cite as http://purl.org/au-research/grants/arc/DP110100221]

**Researchers** Roberts A/Prof Ann; Dr Timothy J Davis

**Brief description** Extending concepts from standard radio-frequency antenna technology down to the nanoscale will open up new applications in fields from biotechnology to telecommunications. This project will embed a light emitting particle in a nanostructured metallic device to produce an ultrabright, directional single-photon source.

**Funding Amount** $490,000

**Funding Scheme** Discovery Projects

I feel obliged to make the following statements as I do not wish to play any role in any misrepresentations:

1) My work was not associated to DP110100221 at any stage. The inclusion of the single line in the manuscript and I quote: "*This research was supported under the Australian Research Council's Discovery Projects funding scheme (Project No. DP110100221)*" was purely due to the request made by my main supervisor (AR) at the time. Her request was neither questioned nor challenged for reasons specified in arXiv:1906.00200v6 [physics.optics].
2) I was neither part of the funding scheme nor its application process, and as such I never benefited from it directly or indirectly.
3) The scheme and its nature were never discussed with myself at any stage.

In fact, I first communicated the idea biperiodic array during the course of my PhD in Mar 2013 to my main supervisor at the time:

**short progress report (Circular Polarisation with Symmetric Cross Cavity Aperture array but Asymmetric lattice)**

**Sead Djalalian-Assl** <s.djalalian-assl@student.unimelb.edu.au> Mon, Mar 11, 2013 at 3:02 AM
To: Ann Roberts <ann.roberts@unimelb.edu.au>

Hi Ann,

Please find attached a short report on what I have been investigating.

Regards,
Amir

**CP_withSymetricCrossCavityArrayButAsymetricLattice.docx**
275K

Snapshots of my CP_withSymetricCrossCavityArrayButAsymetricLattice.docx (page by page) are as follow:

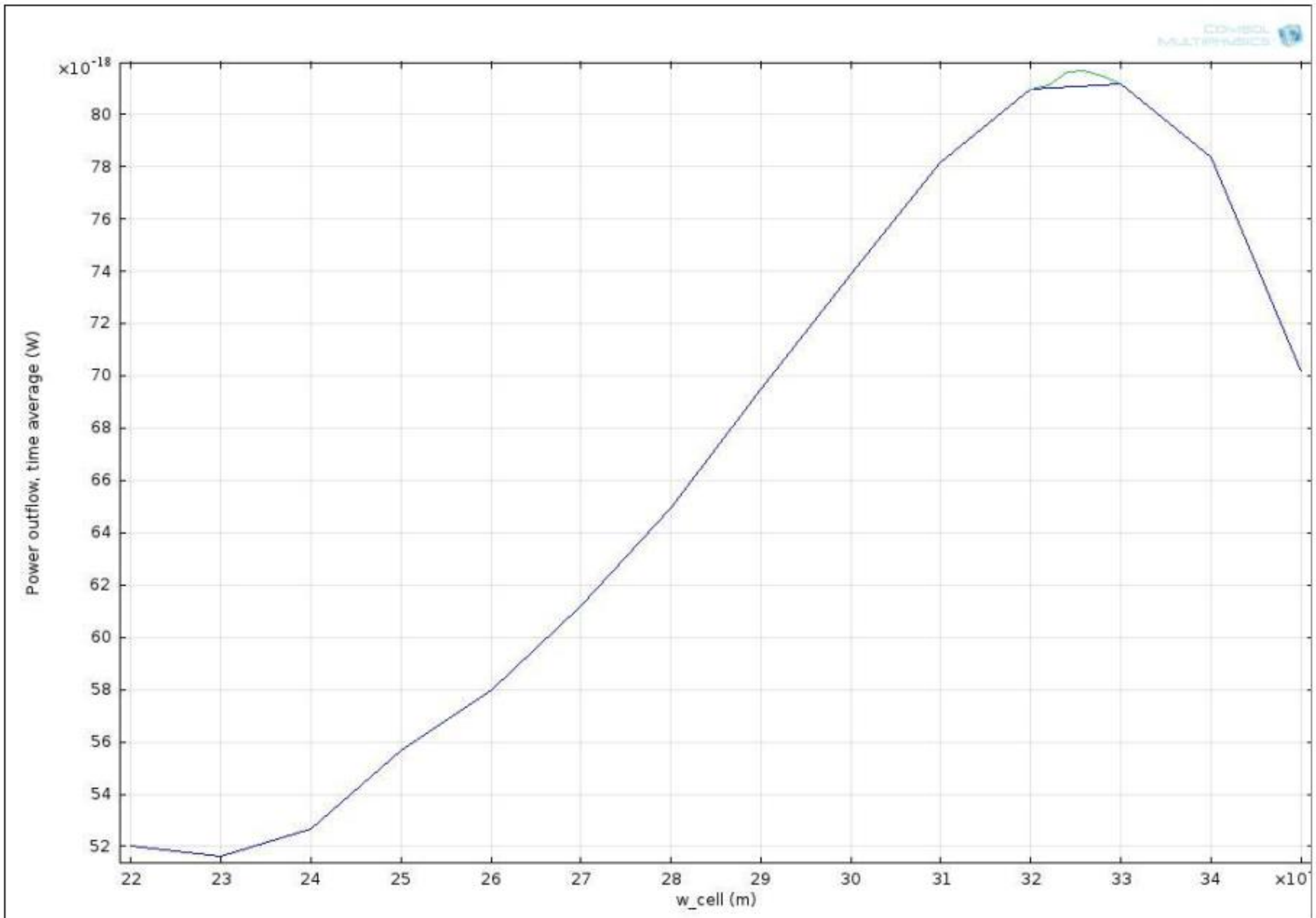

**Figure 1: Transmission spectra of Symmetric Cross Cavity Aperture array with arm-lengths of 170nm but varying in lattice (cell) dimension when incident with 700nm@45°. Width of the cell=Depth of the cell=326nm is a good candidate. Note: film thickness=100nm.**

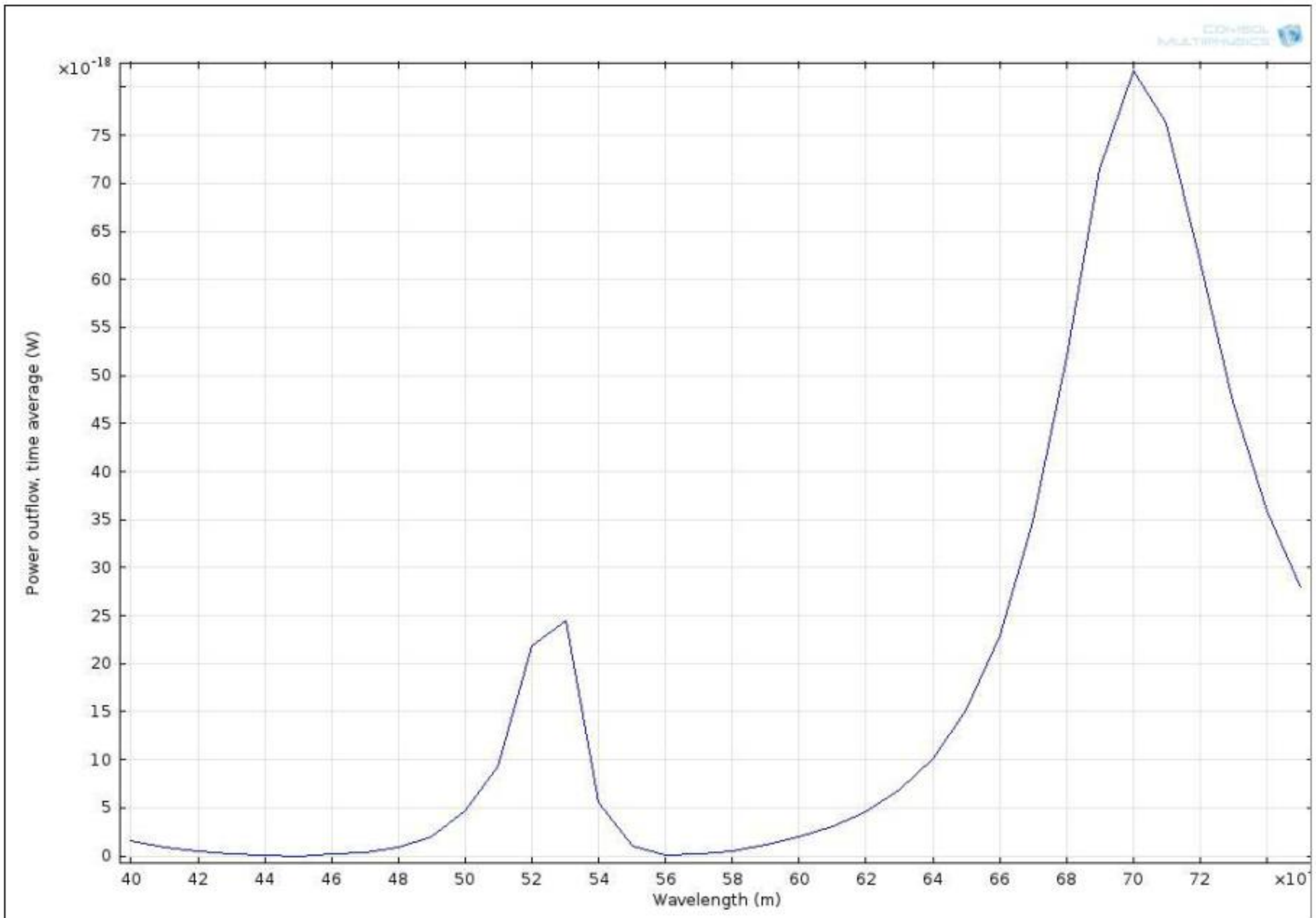

**Figure 2: Transmision spectrum of a Symmetric Cross Cavity Aperture array having armlengths of 170nm and cell sides of 326nm obtained from above. Note the self-consistency resonance at 700nm. Also the resonance at 530nm may be of use in NV deployment.**

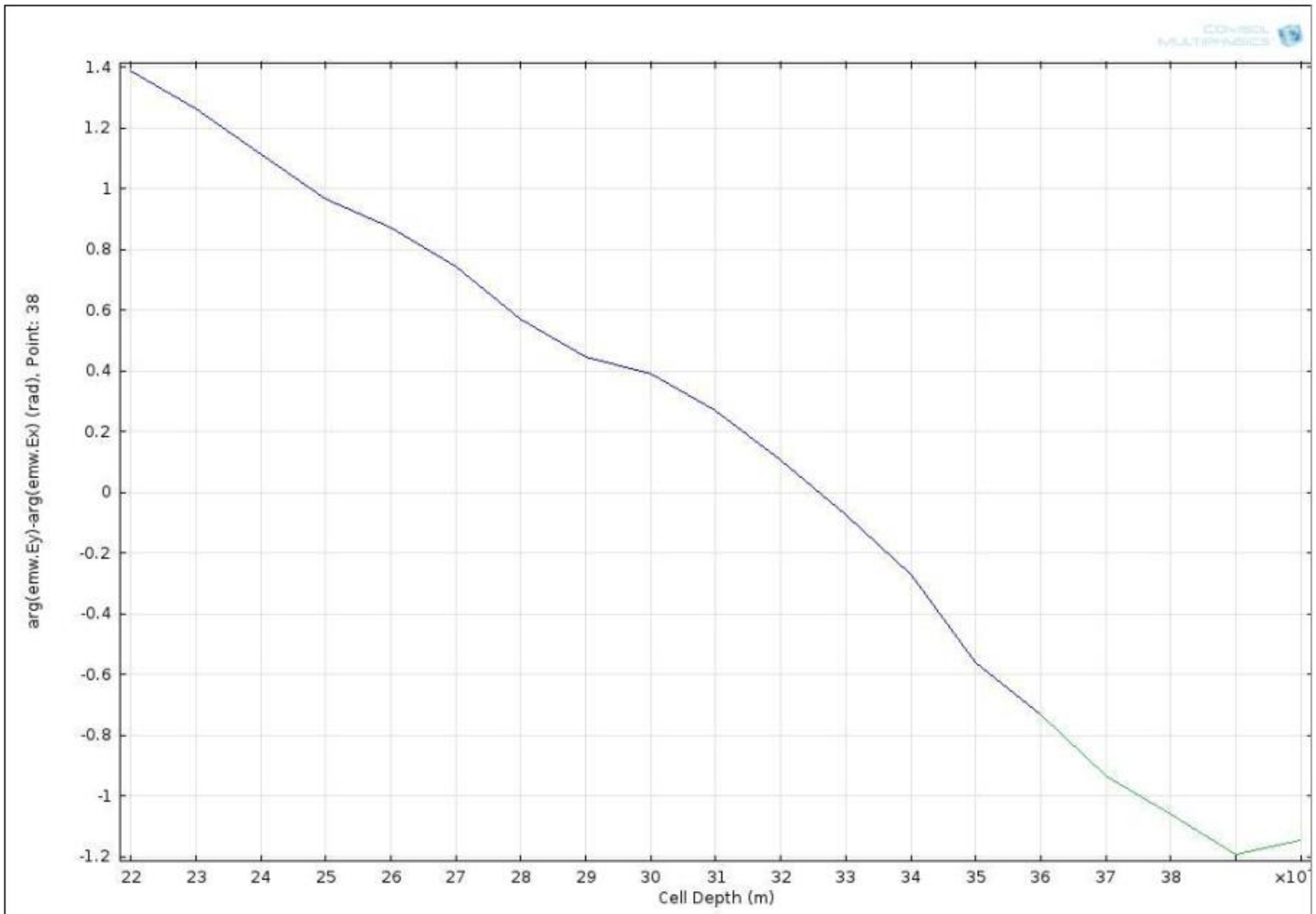

**Figure 3: One side of the cell is set to 326nm and the other varies between 220 and 400nm. Phase difference between the x and y components of the electric field (measured far above the array) vs detuning of one cell side indicates that 90° phase difference is achieved when Width of the cell=226nm and Depth of the cell=362nm is 90 degrees.**

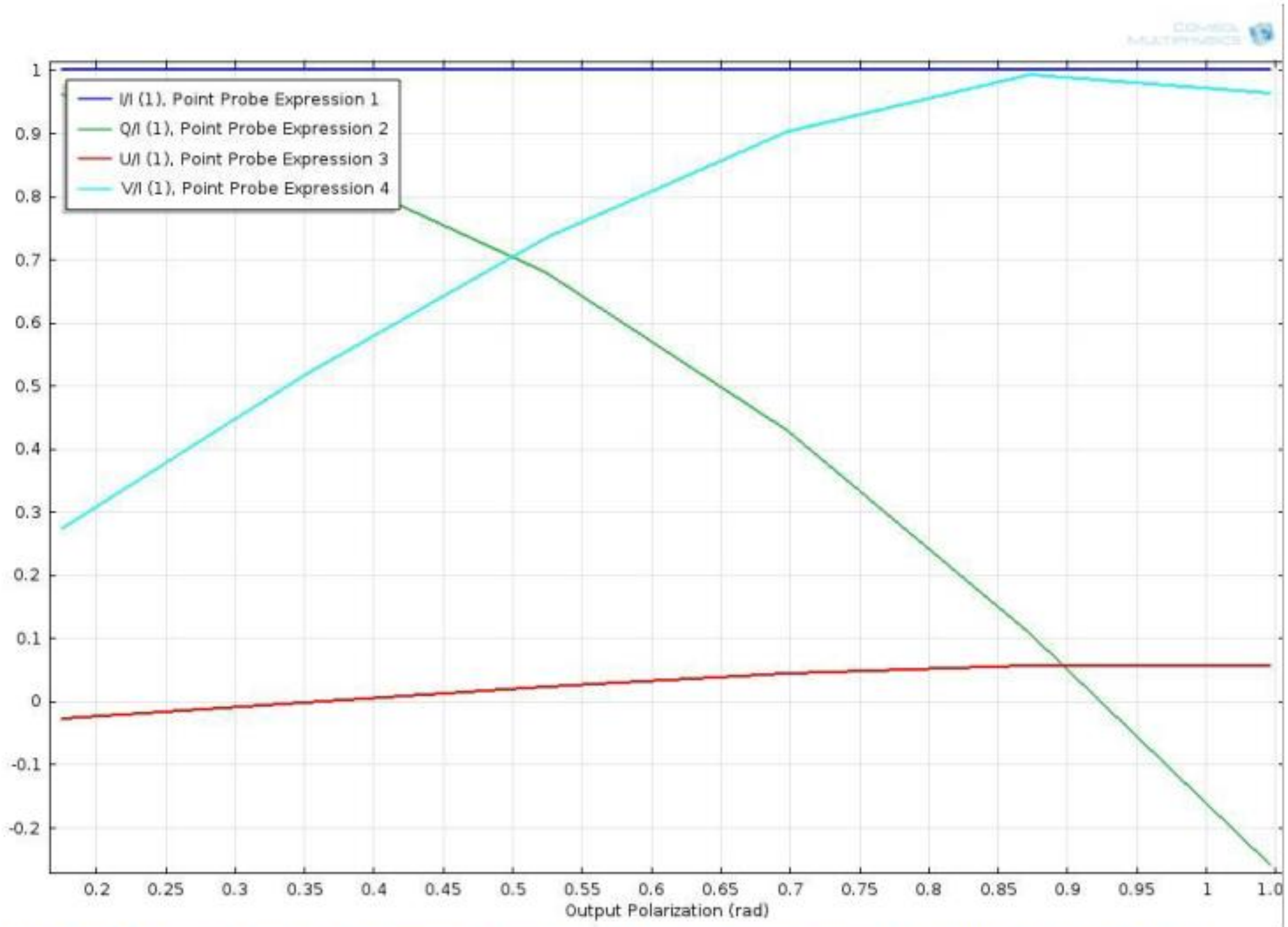

Figure 4: Stokes parameters for the Symmetric Cross Cavity Aperture array with L=170, Cell Width=226 and Cell Depth=362 nm. Polarization of the incident wave is varied in searching for an incident angle that produces a circularly polarized transmitted light.

| I | Q | U | V |
|---|---|---|---|
| 1 | 0.06303 | 0.05701 | 0.99638 |

Table 1: Stokes parameters for transmitted light from the array above when wavelength =700@51.27°

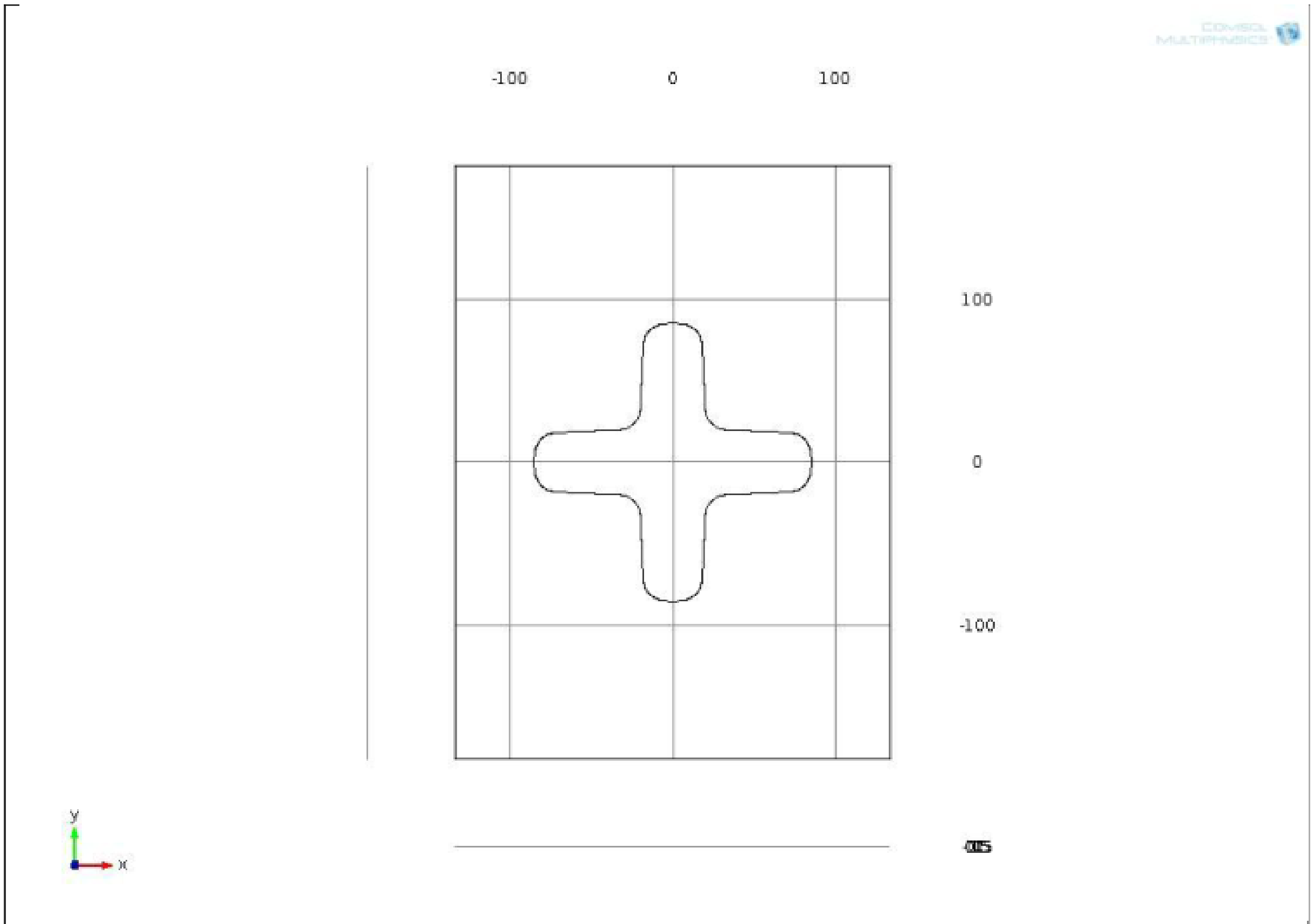

Figure 5: Top view of the final model

I believe that this lattice induced circular polarization can be achieved with most cavities, such as simple holes.

I think this is publishable. What do you think?

---

And the response I received:

**short progress report (Circular Polarisation with Symmetric Cross Cavity Aperture array but Asymmetric lattice)**

**Ann Roberts** <ann.roberts@unimelb.edu.au> Mon, Mar 11, 2013 at 9:53 AM
To: Sead Djalalian-Assl <s.djalalian-assl@student.unimelb.edu.au>
Cc: Ann Roberts <ann.roberts@unimelb.edu.au>

Hi Amir

Thanks for that.

Can you please write some words around this? Does it agree with theory etc? Publishability is going to depend on the story around the measurements, a discussion of novelty and how it fits in with previous work etc. It is possibly publishable somewhere, but write it up (for your thesis at least) and we can decide what to do with it.

Ann

I have elaborated on the gradual development of the idea and its significance in arXiv:1806.00619 [physics.optics].